\documentclass[
aps,prd,showpacs,twocolumn,
superscriptaddress]{revtex4-1}

\usepackage{graphicx}
\usepackage{dcolumn}

\usepackage{amsmath}
\usepackage{amssymb}
\usepackage{bm}
\usepackage{color}
\usepackage[normalem]{ulem}
\usepackage[dvipsnames]{xcolor}
\usepackage{hyperref}
\usepackage{tikz}
\hypersetup{
	colorlinks=true, % false: boxed links; true: colored links
	linkcolor=blue,  % color of internal links
	citecolor=cyan,  % color of links to bibliography
}
\newcommand{\be}{\begin{equation}}
\newcommand{\ee}{\end{equation}}
\newcommand{\bea}{\begin{eqnarray}}
\newcommand{\eea}{\end{eqnarray}}

\begin{document}
	
%%%
\title{On the properties of a deformed extension of the NUT space-time}
%%%%

\author{Bakhtiyor Narzilloev}
\email{nbakhtiyor18@fudan.edu.cn}
	\affiliation{Center for Field Theory and Particle Physics and Department of Physics, Fudan University, 200438 Shanghai, China }
	\affiliation{Ulugh Beg Astronomical Institute, Astronomicheskaya 33, Tashkent 100052, Uzbekistan}

\author{Daniele Malafarina}
	\email{daniele.malafarina@nu.edu.kz}
	\affiliation{Department of Physics, Nazarbayev University, 53 Kabanbay Batyr avenue, 010000 Astana, Kazakhstan }

	\author{Ahmadjon Abdujabbarov}
	\email{ahmadjon@astrin.uz}
    \affiliation{Shanghai Astronomical Observatory, 80 Nandan Road, Shanghai 200030, P. R. China}
\affiliation{Ulugh Beg Astronomical Institute, Astronomicheskaya 33, Tashkent 100052, Uzbekistan}
\affiliation{National University of Uzbekistan, Tashkent 100174, Uzbekistan}
\affiliation{Institute of Nuclear Physics, Ulugbek 1, Tashkent 100214, Uzbekistan}
\affiliation{Tashkent Institute of Irrigation and Agricultural Mechanization Engineers, Kori Niyoziy, 39, Tashkent 100000, Uzbekistan}

	\author{Cosimo Bambi}
	\email{bambi@fudan.edu.cn}
	\affiliation{Center for Field Theory and Particle Physics and Department of Physics, Fudan University, 200438 Shanghai, China }

	\date{\today}

\begin{abstract}

We consider a class of space-times given by a stationary extension of the Zipoy-Voorhees metric that was found by Halilsoy. We show that the solutions do not describe rotating sources but must be interpreted, similarly to the NUT case, as deformed sources endowed with a gravitomagnetic charge. 
We show that the Halilsoy family is directly linked to the NUT space-time, which can be obtained in the limit of vanishing deformations.
We investigate the motion of test particles and photons in this class of space-times, in particular the innermost stable circular orbits and photon capture radius. 
Finally we show that this class of solutions possesses a sub-manifold where closed time-like curves are allowed.

\end{abstract}

\maketitle

\section{Introduction}

Vacuum, exact solutions describing stationary space-times are of great importance in General Relativity since they can describe the field outside a rotating compact source. The most famous of such solutions is the well known Kerr metric which describes the field of a rotating black hole
\cite{kerr63}.

However other solutions do exist and investigating their properties and to what extent they may describe the exterior of physical sources, is important in order to establish the extent to which a solution may be considered physically viable and, as a consequence, whether astrophysical black hole candidates are well described by mathematical black hole solutions
\cite{bh-tests}.
An important class of exact solutions of Einstein's equations that describes deviations from black hole space-times is the so-called Weyl class of static axially symmetric vacuum solutions
\cite{weyl}. 
Given the one-to-one correspondence between metrics belonging to Weyl's class and solutions of Laplace equation in flat two-dimensional space, all static axially symmetric solutions are in principle known 
\cite{hp}.
For example in recent times some attention has been given to the Zipoy-Voorhees (ZV) metric which is a static generalization of the Schwarzschild solution to include higher multipole moments and describes the field outside prolate or oblate spheroids
\cite{zv}.
The properties of the motion of test particles in the ZV space-time and the possibility of testing the geometry from astrophysical observations has been discussed in several articles
\cite{zv2}.
However, astrophysical compact objects typically rotate and therefore it would be more interesting to study the properties of stationary solutions.
This has led many authors in the past to consider `rotating' generalization of static axially symmetric solutions (see for example 
\cite{rotating}). 
In particular, stationary generalizations of the ZV metric have been studied in
\cite{rotating-gamma}.
However, as it turns out, not all such generalizations describe rotating objects.
There exist a number of solution generating techniques that can be adopted in order to obtain a stationary solution from a known static one
\cite{rotating-2}.
Such techniques have been employed in the past to derive a large number of `rotating' generalizations of known solutions. However, the physical properties of these new solutions have rarely been investigated.
One notable exception, i.e. a stationary solution that has been thoroughly investigated, is the so-called Newman-Unti-Tamburino (NUT) space-time \cite{nut}, which, due to its peculiar properties, was at some point referred to as a `counterexample to everything' 
\cite{misner}.
{However, arguments exist in support to the fact that the NUT solution could posses some physical validity and the NUT parameter could in fact be related to angular momentum (see for example
\cite{wu}).}
Similarly to Kerr, the NUT space-time is fully characterized by two parameters, one related to the mass of the source, and the other, the so-called NUT parameter, related to the gravitomagnetic charge of the source. However, and differently from the Kerr case, the NUT parameter does not characterise the rotation of the source. In this respect, the NUT solution constitutes the best example of a metric which may not describe a viable source of the gravitational field but which is still extremely useful to understand the physical properties of exact solutions of Einstein's equations.

In the present article we consider another stationary space-time that was obtained by Halilsoy in 
\cite{halilsoy}
and investigate its properties. The stationary solution was constructed using the ZV space-time as the `seed' metric. 
We show that, contrary to what is claimed in the title of Halilsoy's paper, the line element does not describe a spinning massive source. On the contrary, the `spin' parameter in the solution behaves similarly to the NUT parameter and we show explicitly that the metric in question reduces to the NUT space-time in the case of vanishing deformations.

The paper is organised as follows: In section \ref{2} we outline the basic features of the stationary extension of the ZV space-time, show its relation with the NUT space-time and discuss its properties as the exterior of a gravitating compact source. Section \ref{3} is devoted to the study of the motion of test particles in the stationary ZV geometry, with particular emphasis on its relation to the Schwarzschild, Kerr and NUT space-times. Finally in section \ref{4} we briefly discuss the results and their implications for astrophysical black holes and exotic compact objects.

Throughout the paper we make use of geometrized units setting $G=c=1$.

\section{Stationary Zipoy-Voorhees metric}\label{2}

The most general stationary and axially symmetric vacuum space-time in Weyl's cylindrical coordinates $\{t,\rho,z,\phi\}$ has the following form
\be\label{line1}
ds^2=-e^{2 \psi} (dt-\omega d\phi)^2+e^{-2\psi}[e^{2\lambda} (d\rho^2+dz^2)+\rho^2 d\phi^2]\, ,
\ee
where $\psi=\psi(\rho,z)$, $\lambda=\lambda(\rho,z)$ and $\omega=\omega(\rho,z)$ are determined from Einstein's equations
\cite{weyl}.

Specific solutions of this class can be obtained from known static axially-symmetric solutions through a variety of procedures 
\cite{rotating-2}.
However, it is well known that not all stationary solutions of the above class describe the exterior gravitational field of a rotating object. The most famous of such examples is the Newmann-Unti-Tamburino (NUT) metric
\cite{nut}.

The NUT space-time belongs to the above class and its metric functions are given by
\bea 
e^{2 \psi}&=& \frac{(r_++r_-)^2-4(M^2+l^2)}{(r_++r_-+2M)^2+4l^2}\, ,\\
e^{2\lambda}&=& \frac{(r_++r_-)^2-4(M^2+l^2)}{4r_+r_-}\, ,\\
\omega&=&\frac{l(r_+-r_-)}{\sqrt{M^2+l^2}}\, ,
\eea 
where $M$ and $l$ are positive parameters and
\be 
r_{\pm}^2=\rho^2+(z\pm\sqrt{M^2+l^2})^2\, .
\ee 
In the case $l=0$ the metric becomes static and it reduces to the Schwarzschild geometry. 
From the asymptotic expansion of the metric the parameter $M$ is immediately interpreted as the gravitational mass of the source. 
However, the interpretation of the NUT solution and of the NUT parameter $l$ is less trivial and it has attracted lots of attention over the years. 
In 
\cite{bonnor}
the solution was interpreted as representing the exterior field of a mass located at the origin together with a semi-infinite massless source of angular momentum.
On the other hand in
\cite{manko}
the metric was interpreted as describing the exterior field of 
two counter-rotating semi infinite rods of negative mass, separated by a static rod of finite length and positive mass.
In 
\cite{LB} 
the parameter was interpreted as a gravitomagnetic charge bestowed upon the central mass. Conversely in 
\cite{al-badawi} 
the parameter was interpreted as a property of the surrounding space-time.
{Another direction towards the interpretation of the NUT space-time comes from the analysis of its thermodynamic properties. In \cite{wu} it was shown that if the NUT parameter is interpreted as possessing simultaneously rotational and electromagnetic features then the thermodynamic equivalent of Bekenstein-Smarr formula follows naturally.}

Another famous solution of the above class is the ZV space-time describing the gravitational field outside a static deformed object \cite{zv}. The metric functions are given by
\bea 
e^{2 \psi}&=& \left(\frac{R_++R_--2m}{R_++R_-+2m}\right)^\gamma \, ,\\
e^{2\lambda}&=& \left(\frac{(R_++R_--2m)(R_++R_-+2m)}{4R_+R_-}\right)^{\gamma^2}\, ,\\
\omega&=&0\, ,
\eea 
where $m$ and $\gamma$ are positive parameters and
\be 
R_{\pm}^2=\rho^2+(z\pm m)^2\, .
\ee 
The space-time reduces to Schwarzschild for $\gamma=1$ and the parameter $\gamma$ is easily interpreted as a deformation parameter. The total gravitational mass of the source is $M=m\gamma$ and from the computation of the quadrupole moment $Q=\gamma m^3(1-\gamma^2)/3$ one can see that values of $\gamma<1$ ($\gamma>1$) correspond to prolate (oblate) deformations.
The properties of the ZV space-time were studied in \cite{zv2}, while interior solutions for the ZV metric were obtained in \cite{gamma}. 

A stationary generalization of the ZV metric was obtained by Halilsoy in 
\cite{halilsoy}. 
If we perform a change of coordinates from cylindrical to prolate spheroidal coordinates $\{t,x,y, \phi\}$ given by
\bea 
\rho&=& \kappa \sqrt{x^2-1}\sqrt{1-y^2}\, ,\\
z&=&\kappa xy\, ,
\eea 
with $\kappa=m/\gamma$, the line element \eqref{line1} takes the form
\begin{eqnarray} \label{line}
ds^2&=&-e^{2 \psi} dt^2+2\omega e^{2 \psi}dtd\phi\\ \nonumber 
&&+\frac{ m^2 (x^2-y^2)  e^{2\lambda-2 \psi} }{\gamma ^2} \left(\frac{dx^2}{x^2-1}+\frac{dy^2}{1-y^2}\right)
\\\nonumber
  &&+ \left(\frac{m^2
   \left(x^2-1\right) \left(1-y^2\right) e^{-2 \psi}}{\gamma ^2}-\omega^2e^{2 \psi}\right)d\phi^2\ ,
\end{eqnarray}
and the metric functions can be given as
\begin{eqnarray}\nonumber
 e^{-2 \psi} &=& \frac{1}{2} \left(\frac{x+1}{x-1}\right)^{\gamma } \left[1+p+(1-p) \left(\frac{x-1}{x+1}\right)^{2 \gamma }\right]\ ,
    \\\nonumber
\\
e^{2\lambda} &=& \left(\frac{x^2-1}{x^2-y^2}\right)^{\gamma ^2}\ ,
\\\nonumber
    \\\nonumber
\omega &=& -2 m q y\ .
\end{eqnarray}
The line element depends on three parameters $m, \gamma$ and $q$ with $p$ given by $p^2+q^2=1$. For $q=0$ one retrieves the ZV space-time so that for $q=0$ (i.e. $p=1$) and $\gamma=1$ one obtains the Schwarzschild metric. Similarly to the space-times discussed above the parameter $m$ is related to the gravitational mass of the source, $\gamma$ is related to the deformation of the source while the parameter $q$ describes the departure from the static solution.
However, one needs to be careful when interpreting $q$ as a rotation parameter similar to the angular momentum in Kerr's solution, since a stationary space-time may be obtained also by means of introducing a NUT-like parameter. Indeed, this turns out to be the case here.

In order to relate the above line element to the Schwarzschild and NUT space-times it is useful to make use of Erez-Rosen coordinates $\{t,r,\theta,\phi\}$ using the transformations
\begin{eqnarray}
\rho^2&=&r^2\left(1-\frac{2m}{r}\right)\sin^2\theta \ ,\\
z&=&(r-m)cos\theta\ ,\\\nonumber
\end{eqnarray}
or equivalently
\begin{eqnarray}
x&=&\frac{r}{m}-1 \ ,\\
y&=&cos\theta\ ,\\\nonumber
\end{eqnarray}
so that the line element \eqref{line1} takes the form
\begin{eqnarray} \label{line}
ds^2&=&-e^{2 \psi}dt^2 + \frac{ e^{2\lambda-2 \psi} \Sigma}{\Delta}dr^2 + e^{2\lambda-2 \psi}  \Sigma r^2 d\theta^2 
\\\nonumber
&&+ \left(e^{-2 \psi} \Delta r^2 \sin^2\theta-\omega^2e^{2 \psi}\right) d\phi^2 +2 \omega e^{2 \psi} dt d\phi,
\end{eqnarray}
and the metric functions become
\begin{eqnarray}
e^{-2 \psi}&=& \frac{1}{2}\Bigg[\Delta^\gamma+\frac{1}{\Delta^\gamma}\Bigg]-\frac{p}{2} \Bigg[\Delta^\gamma-\frac{1}{\Delta^\gamma}\Bigg]\ ,\\
e^{2\lambda}&=& \frac{\Delta^{\gamma^2}}{\Sigma^{\gamma^2}} \ ,\\
\omega&=& -2 m \gamma q\, cos\theta \ ,
\end{eqnarray}
with
\begin{eqnarray}
\Delta&=&1-\frac{2 m}{r} \ ,\\
\Sigma&=&1-\frac{2 m}{r}+\frac{m^2}{r^2} sin^2\theta \ ,\\
p&=&\sqrt{1-q^2} \ .
\end{eqnarray}

From the above changes of coordinates we see that in general the coordinates have range $r\in(2m,+\infty)$, $\theta\in(0,\pi)$ for Erez-Rosen coordinates and $x\in[1,+\infty)$ and $y\in[-1,1]$ in prolate spheroidal coordinates. It can be easily checked that setting $q=0$ one recovers the ZV metric in the usual form given by Erez-Rosen coordinates. However, as mentioned before, for $q\neq 0$ and $\gamma=1$ one does not retrieve the Kerr solution. In fact the line element for the case $\gamma=1$ becomes
\begin{eqnarray} \label{line-gamma=1}
ds^2&=&-\frac{\Delta}{F}(dt-\omega d\phi)^2+F\left(\frac{dr^2}{\Delta}+r^2d\Omega^2\right),
\end{eqnarray}
with
\be 
F=1+(1-p)\frac{2m}{r}\left(\frac{m}{r}-1\right)\, .
\ee 
It is not difficult to see that for $p=1$ (i.e. $q=0$) we retrieve the Schwarzschild solution in Schwarzschild coordinates. 
However, to retrieve the NUT solution in the usual form when $\gamma=1$ and $q\neq 0$ one needs to perform the change of coordinates
\be \label{coord}
\bar{r}=r-m(1-p)\, ,
\ee 
and identify the parameters from
\bea \label{l}
l&=&mq \, , \\ \label{m}
M&=&mp \, .
\eea
Then the line element \eqref{line-gamma=1} takes the familiar form
\be 
ds^2=-\frac{\delta}{\sigma}(dt-\omega d\phi)^2+\sigma\left(\frac{d\bar{r}^2}{\delta}+d\Omega^2\right),
\ee 
with
\bea 
\delta&=&\bar{r}^2-2M\bar{r}-l^2 \, , \\
\sigma&=&\bar{r}^2+l^2\, ,\\
\omega&=&-2l\cos\theta\, .
\eea 

{This re-interpretation of the NUT parameter $l$ is consistent with the ideas put forward in \cite{manko} and \cite{wu}. In fact, the transformation \eqref{coord} shifts the inner and outer horizons to $\tilde{r}_-=0$ and $\tilde{r}_+=2m$, respectively. A detailed analysis of the thermodunamical properties of the NUT solution is beyond the scope of this work. However it is worth mentioning that from the above identification of the parameters one find the formula for the surface gravity $\kappa$ at the outer horizon as $\kappa=1/(2m(p+1))$. Similarly, the area formula for the outer horizon becomes $A=8\pi m^2(p+1)$ which in turn allows to formulate the first law.}

\begin{figure*}[t!]
\begin{center}
\includegraphics[width=0.485\linewidth]{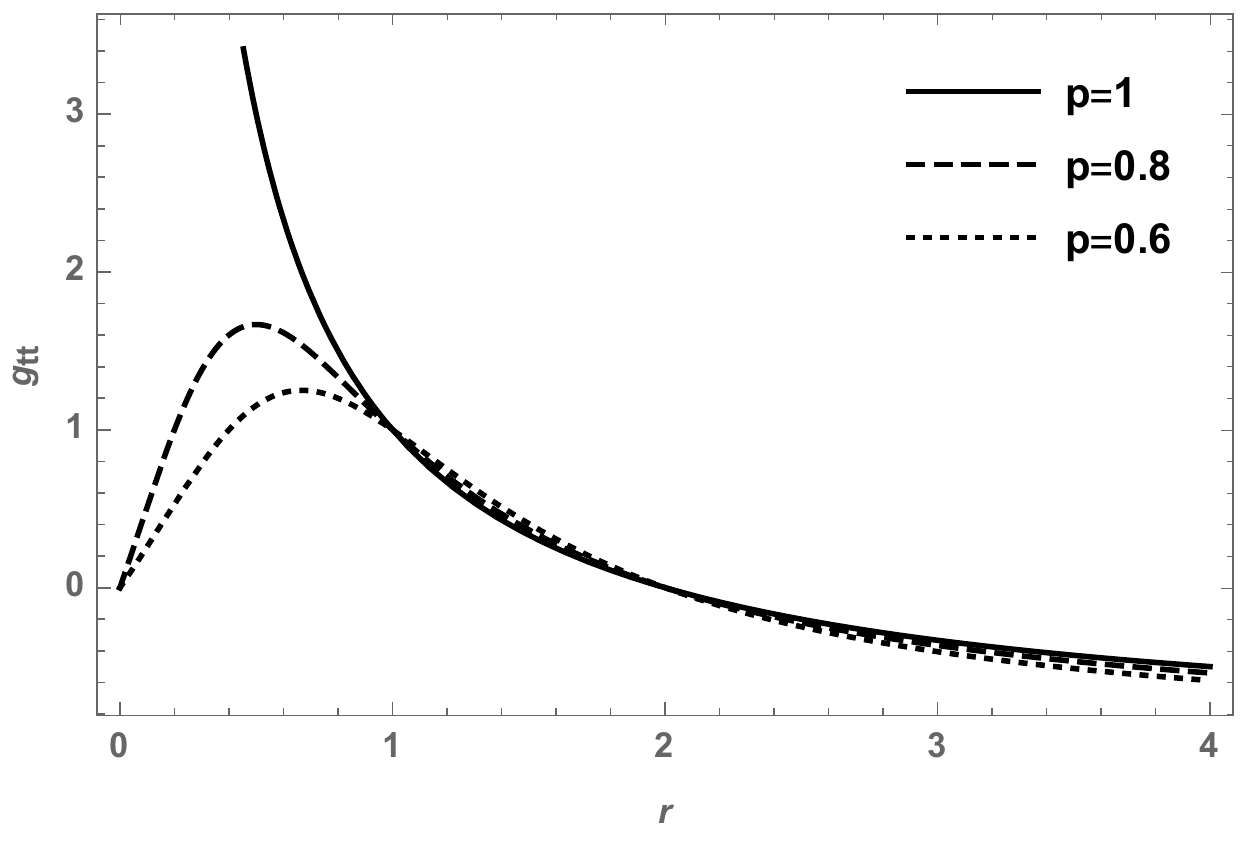}
\includegraphics[width=0.5\linewidth]{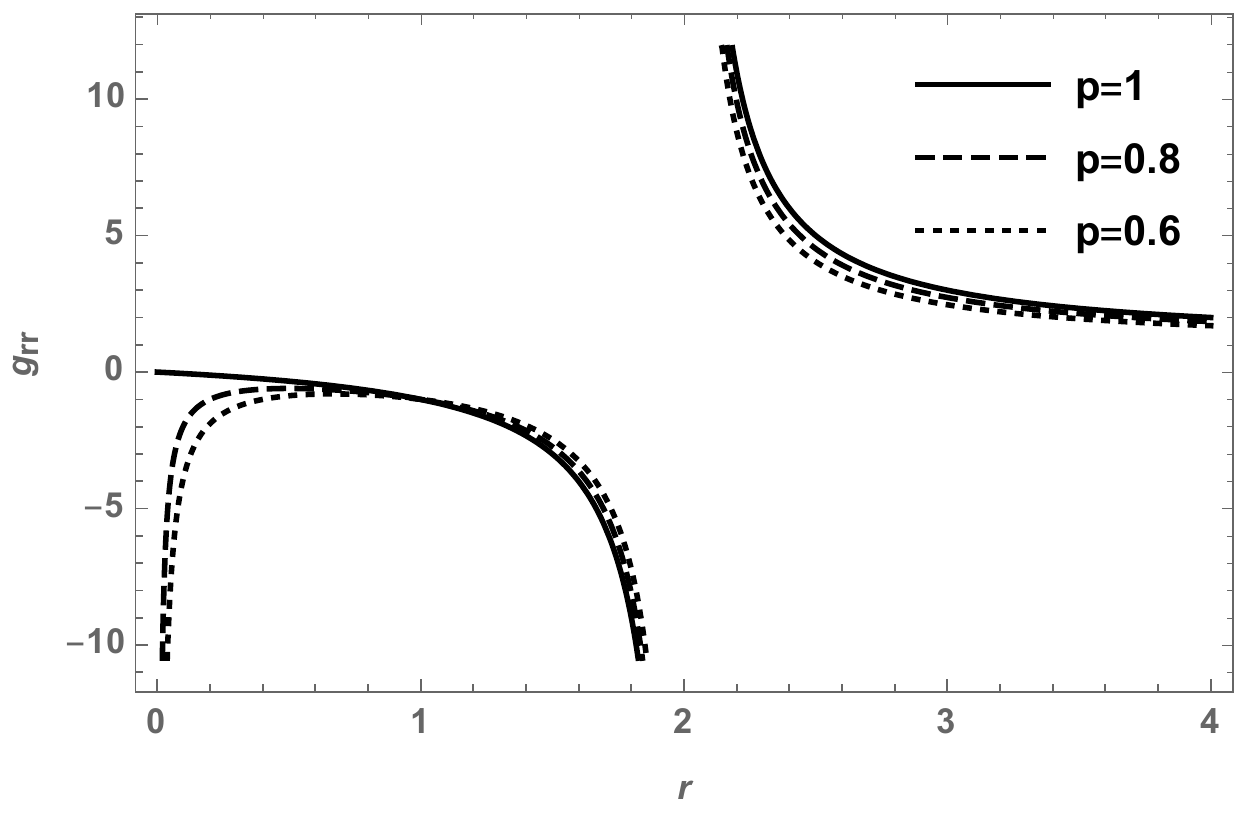}
\end{center}
\caption{The radial dependence of the metric components $g_{tt}$ (left panel) and $g_{rr}$ (right panel) for fixed value of the mass parameter $m$, in the NUT space-time (i.e. for $\gamma=1$) with line element \eqref{line-gamma=1} and different values of the `quasi-NUT' parameter $q$.
\label{gttgrr}}
\end{figure*}

It is easy to see that the metric (\ref{line}) reduces to Minkowski space-time for any value of $\gamma$ when $m=0$.
This also shows that Minkowski can be obtained as the zero mass limit for the NUT metric, if we set $m=0$, instead of the usually discussed limit 
\begin{eqnarray}\nonumber
ds^2&=& -H(r) dt^2 + H(r)^{-1} dr^2 + (l^2+r^2) d\theta^2
\\
&&+ \left[\left(l^2+r^2\right)\sin ^2\theta  -4 l^2 H(r) \cos ^2\theta \right]  d\phi^2 \\\nonumber
 &&-4  l H(r) \cos \theta  dt d\phi ,
\end{eqnarray}
with $H(r)=(r^2-l^2)/(r^2+l^2)$, which is obtained by setting $M=0$ while keeping $l\neq 0$. 
Also it is worth noticing that the coordinate transformation \eqref{coord} maps the horizons of the NUT line element $\bar{r}_h=M\pm\sqrt{M^2+l^2}$ into $r_i=0$ and $r_o=2m$.

In the case of small deviation from spherical symmetry ($\delta=\gamma-1<<1$) and small deviations from staticity ($q\simeq 0$), the components of the metric (\ref{line}) can be expanded as
\begin{eqnarray}\nonumber
g_{tt}&=&-\Delta \left[1+\delta \log \Delta\right]\ ,\\\nonumber
g_{rr}&=&\frac{1}{\Delta}\left[1+\delta \log \left(\frac{\Delta}{\Sigma^2}\right)\right]\ ,\\
g_{\theta\theta}&=&r^2 \left[1+\delta \log \left(\frac{\Delta}{\Sigma^2}\right) \right]\ ,\\\nonumber
g_{\phi\phi}&=&r^2 (1-\delta \log \Delta ) \sin ^2\theta \ ,\\\nonumber
g_{t\phi}&=&-2 \Delta  m q (1+\delta+\delta \log \Delta )\cos \theta\ .
\end{eqnarray}

If we look at the asymptotic properties of the metric \eqref{line} we find the following expressions
\begin{eqnarray}\label{g00}
g_{tt}&=&-\left(1-\frac{2 \gamma m p}{r}\right)+O\left(\frac{1}{r^2}\right)\ ,\\
g_{t\phi}&=&-2 \gamma m q \left(1-\frac{2 \gamma m p}{r}\right)  cos\theta +O\left(\frac{1}{r^2}\right)\ .
\end{eqnarray}
From equation (\ref{g00}) one can see that the gravitational mass of the source is given by $M=\gamma m p$. However, from the expansion of $g_{t\phi}$ we see how the parameter $q$ 
is not related to the angular momentum of the central object, as, otherwise, it should appear as $q/r$ at the leading order in $g_{t\phi}$.

The above statement can be made more precise by evaluating the Komar integrals in the NUT limit, i.e. in the case $\gamma=1$. 
The Komar angular momentum $J_K$ is given by (see \cite{kent}):
\begin{eqnarray}
J_{K}=-\frac{1}{4} \int^{\pi}_{0} (\phi^{\alpha;\beta}n_{[\alpha}r_{\beta]}\sqrt{g_{\theta\theta}g_{\phi\phi}})|_{r\rightarrow\infty}d\theta=0\, ,
\end{eqnarray}
where $\phi^{\alpha}$ is a space-like Killing vector, while $n_{\alpha}$ and $ r_{\alpha}$ are the unit vectors normal to the $t=const$ and $r=const$ hypersurfaces, respectively.
In the Kerr space-time the Komar angular momentum is $J=ma$. On the other hand,
in the NUT space-time the Komar angular momentum is equal to zero.
The fact that the integral above for the metric \eqref{line-gamma=1} vanishes supports the idea that the parameter $q$ does not describe rotation but is better interpreted as some NUT-like charge that we shall call for simplicity `quasi-NUT' parameter, which is directly related to the NUT parameter in the case $\gamma=1$ from equation \eqref{l}.
{However it must be noted that the vanishing of the Komar angular momentum may be interpreted in different ways. In fact, it has been argued that an alternative interpretation of the NUT solution may be given as the exterior field of two counter-rotating semi infinite rods of negative mass, separated by a static rod of finite length and positive mass, all placed along the symmetry axis \cite{bonnor}. This argument supports the interpretation of the NUT parameter as physically valid and related to angular momentum.}

Similarly one can calculate the Komar mass $M_K$ of the source for the metric \eqref{line-gamma=1}. This is given by
\begin{eqnarray}
M_{K}&=&\frac{1}{4} \int^{\pi}_{0} (t^{\alpha;\beta}n_{[\alpha}r_{\beta]}\sqrt{g_{\theta\theta}g_{\phi\phi}})|_{r\rightarrow\infty}
d\theta=mp\ .
\end{eqnarray}
As expected we see that the result coincides with the NUT mass parameter. Also, the Komar mass evaluated in the general case gives $M_K=m\gamma p$ in agreement with the result obtained from the asymptotic expansion in equation \eqref{g00}. Similarly to the ZV metric the Halilsoy's metric exhibits a curvature singularity for $r=2m$ for all values of $\gamma\neq 1$, as can be seen from evaluation of the Kretschmann scalar $K=R_{\alpha\beta\gamma\sigma}R^{\alpha\beta\gamma\sigma}$,  which characterizes the space-time curvature.
On the other hand, one can check that for $\gamma=1$, similarly to Schwarzschild, $g_{tt}$ changes sign at $r=2m$ while $g_{rr}$ diverges, as can be seen from Fig.~\ref{gttgrr}. Evaluation of the Kretschmann scalar in this case shows that $r=2m$ is regular and coincides with the horizon of the NUT space-time. 
It is easy to check that for $\gamma=1$ the Kretschmann scalar $K$ does not diverge for any value of $q\neq 0$, thus showing that the singularity given by the divergence of $g_{rr}$ at $r=0$ is also a coordinate singularity and the space-time can be extended to the range $r\in(-\infty,+\infty)$. In this case $K$ is given by 
\begin{eqnarray}
    &K&=\frac{48 m^2h(r)}{\left[-2 m^2 (p-1)+2 m (p-1) r+r^2\right]^6} \, ,
\end{eqnarray}
with
\begin{eqnarray}\nonumber
h(r)&=& 8 m^6 (p-1)^3 p+24 m^5 (p-1)^3 r+\left(2
   p^2-1\right) r^6  \\\nonumber
   && +60 m^4 (p-1)^2 r^2-40 m^3 (p-1)^2 r^3\\\nonumber
  && +30 m^2 (p-1) p r^4 +6 m \left(-2 p^2+p+1\right) r^5 .
\end{eqnarray}
Also, it is easy to notice that in the coordinates used in \eqref{line-gamma=1} for $p\neq 1$ we have $K(0)=-192m^6p(p-1)^2<0$, thus suggesting that repulsive effects appear in the vicinity of the center of the space-time.

Further we can investigate the departure from spherical symmetry of the source of the metric \eqref{line-gamma=1} by calculating the area of the surfaces with constant $r$ surrounding the source.
Therefore we must use the three-dimensional line element given by \cite{nbm0}
\begin{eqnarray}
dl^2=\gamma_{ij} dx^{i} dx^{j} \ ,
\end{eqnarray}
where
\begin{eqnarray}
\gamma_{ij}=g_{ij}-\frac{g_{ti}g_{tj}}{g_{tt}} \ .
\end{eqnarray}
Then the surface becomes
\begin{eqnarray}
\mathcal{S}(r)=4 \pi  r^2 \left[1+\frac{2 m }{r}(p-1)-\frac{2 m^2 }{r^2}(p-1)\right]\ .
\end{eqnarray}
It is clear that in the static case (i.e. $p=1$) we recover the revolution surfaces in the Schwarzschild geometry, $\mathcal{S}_{\rm Sch}=4 \pi r^2$, while for $p\neq 1$ we obtain $\mathcal{S}<\mathcal{S}_{\rm Sch}$. Also for $p\neq 1$ there is a limiting radius given by
\be
r_0=m(\sqrt{(p-1)^2-2(p-1)}-p+1)\, ,
\ee
at which $\mathcal{S}(r_0)=0$. 

\begin{figure}
	\begin{center}
		\includegraphics[scale=0.6]{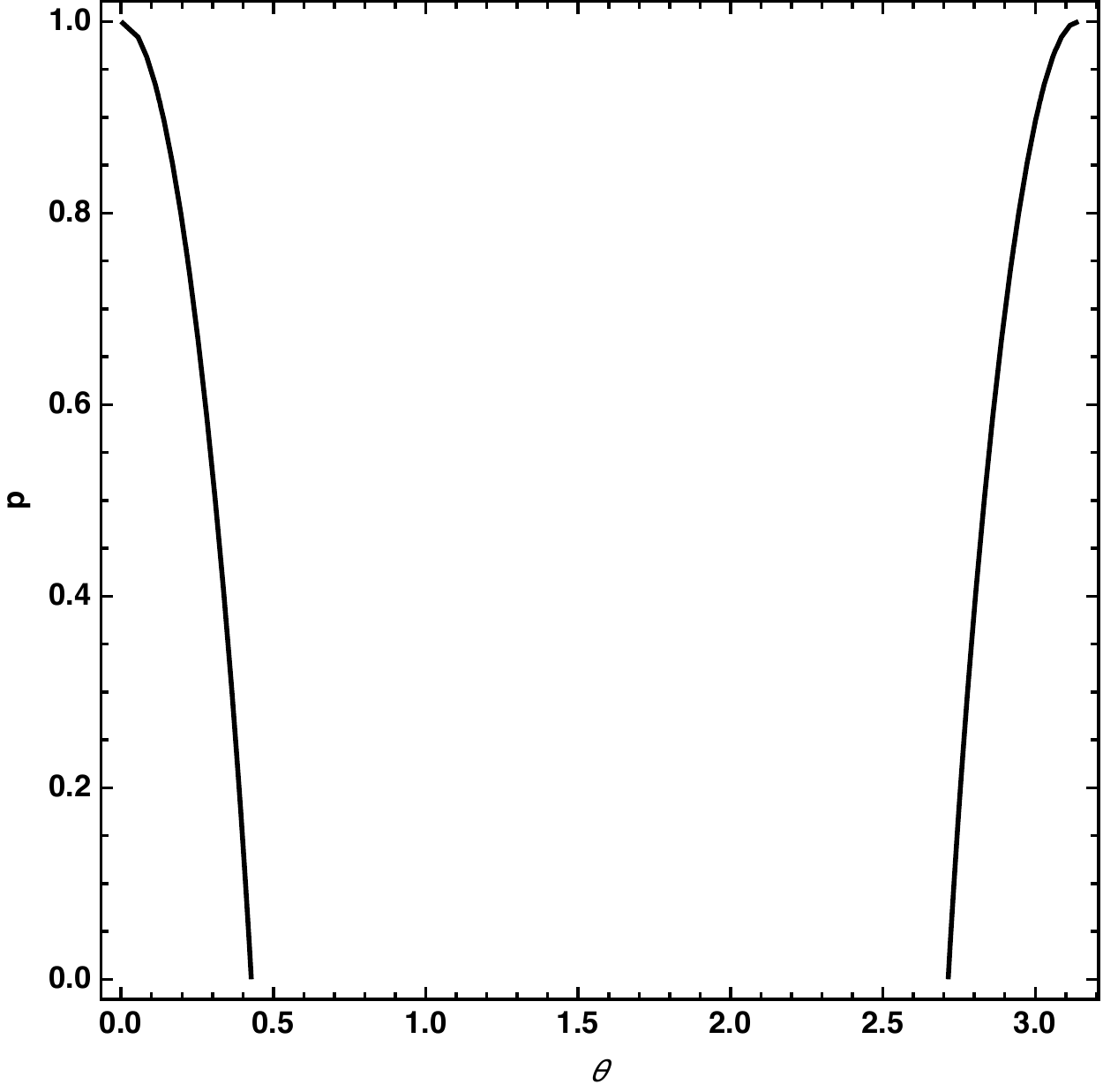}
		\caption{The line for which $g_{\phi\phi}$ is equal to zero for the stationary ZV space-time. In the outer part of the line $g_{\phi\phi}$ is positive and geodesics with constant $t$, $r$ and $\theta$ are space-like. In the inner part of the line $g_{\phi\phi}$ becomes negative corresponding to appearance of closed time-like curves. The closed time-like curves of the NUT metric, $q=0$, in the coordinates of line element \eqref{line-gamma=1} are shifted to negative values of $r$. \label{g3}}
	\end{center}
\end{figure}

Finally, when thinking about the physical validity of the line element \eqref{line-gamma=1} it is worth checking whether pathologies such as closed time-like curves do appear.
One simple way to do so is to study the properties of geodesics with constant values of the coordinates $t$, $r$ and $\theta$. The interval reduces to
\begin{eqnarray}
ds^{2}=g_{\phi\phi} d\phi^{2}\, ,
\end{eqnarray}
and it is space-like for $g_{\phi\phi}>0$. It is easy to see that for $\gamma\neq 1$ and $p\neq 1$ there exist regions where $g_{\phi\phi}$ turns from space-like to time-like thus showing that closed time-like curves can appear in those cases. For any fixed value of $r$ one may find the regions where closed time-like curve are allowed by finding the zeroes of $g_{\phi\phi}$ as a function of $p$ and $\theta$ (see Fig.~\ref{g3}). On the other hand, for $q=0$, i.e. the ZV metric, and for $\gamma=1$, i.e. the NUT metric, we see that $g_{\phi\phi}$ remains positive everywhere for positive values of $r$.

\section{particle motion}\label{3}

\begin{figure*}[t!]
\begin{center}
\includegraphics[width=0.49\linewidth]{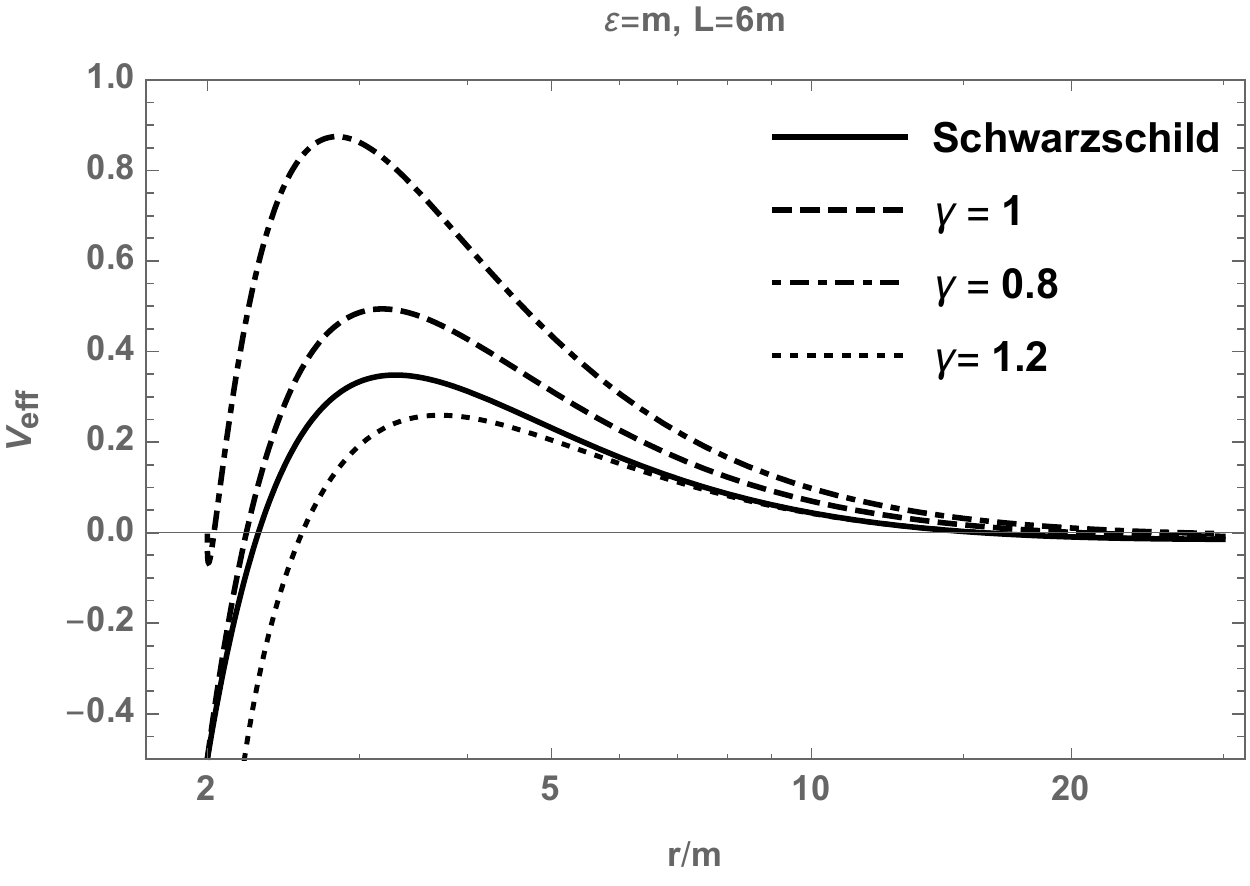}
\includegraphics[width=0.49\linewidth]{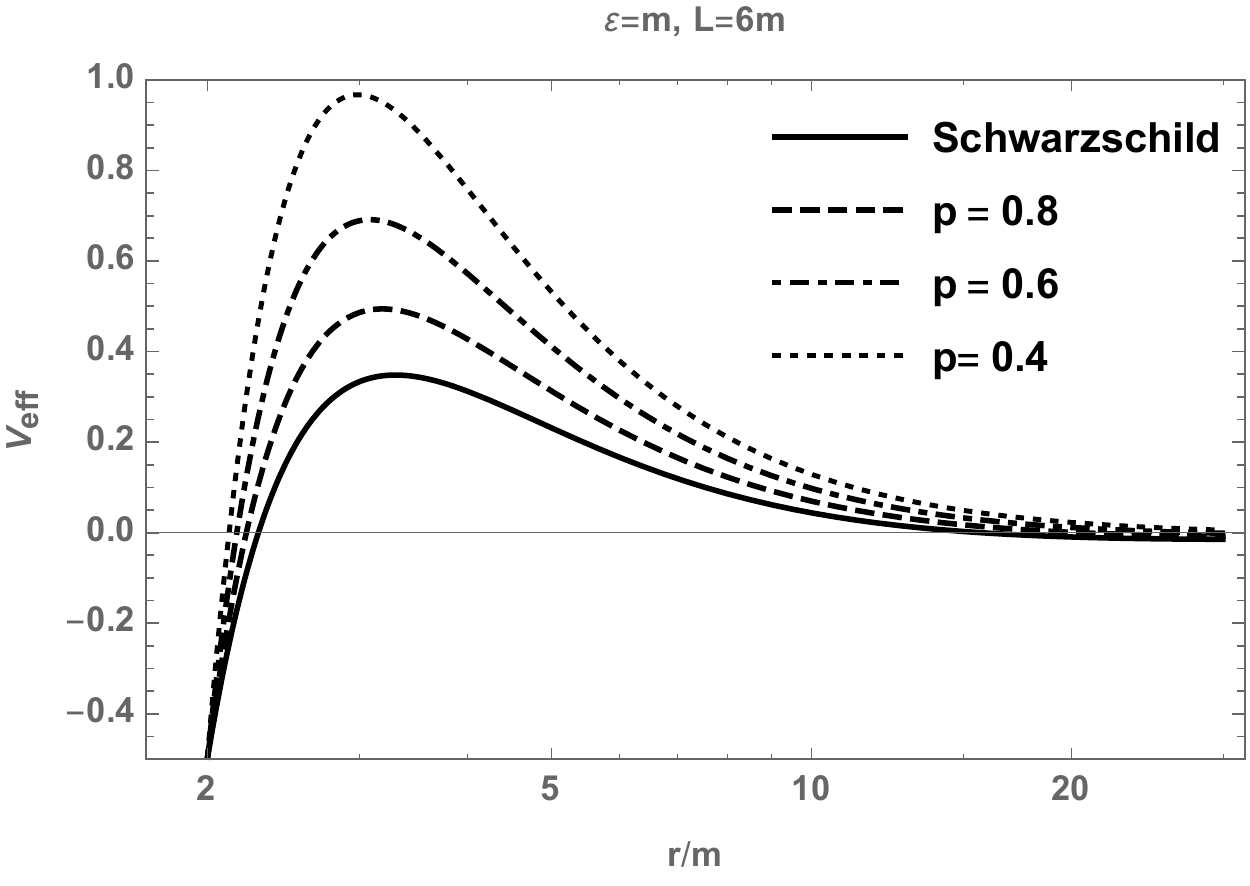}
\end{center}
\caption{The radial dependence of the effective potential $V_{eff}(r)$ for massive test
particles in the equatorial plane of the stationary ZV space-time as compared to the effective potential for Schwarzschild.  In the left panel $V_{eff}(r)$ is plotted for $p=0.8$ and various values of $\gamma$. In the right panel $V_{eff}(r)$ is plotted for $\gamma=0.8$ and various values of $p$. As expected, notable departures from the Schwarzschild case appear for $\gamma<1$ and $p\simeq 0$.
\label{veff_1}}
\end{figure*}

In order to investigate the viability of the line element as a possible exterior gravitational field of a physical source it is worth looking at the motion of test particles.
Therefore we shall now outline the formalism to describe test particle motion in the metric given by equation (\ref{line}). Since the metric does not depend on the coordinates $t$ and $\phi$ we have the two usual time-like and space-like Killing vector fields associated with time translations and rotations. Therefore the action $S$ for the particle can be written in the usual form
\begin{eqnarray}
S= -\varepsilon t + \mathcal{L} \phi + S(r,\theta)\ ,
\end{eqnarray}
where $\varepsilon$ and $\mathcal{L}$ are the constants of motion associated with the two Killing vectors and representing the energy and angular momentum per unit mass of the test particle. Using Hamilton-Jacobi equation
\begin{equation}\label{HJ}
g^{\alpha\beta} \frac{\partial S}{\partial x^{\alpha}} \frac{\partial S}{\partial x^{\beta}}=-1\ ,
\end{equation}
we find the effective potential $V_{eff}(r)$ for test particles in the equatorial plane ($\theta=\pi/2$). The radial component of the four velocity of the particle reads
\begin{eqnarray}\label{f}
\dot{r}^2=f(r)=\varepsilon^2-1-2 V_{eff}(r)\ ,
\end{eqnarray}
and the effective potential takes the form
\begin{eqnarray}\nonumber
V_{eff}(r)&=&\frac{1}{2} \left[\frac{r \left( \varepsilon^2 e^{-2 \psi}-1\right) (2 m-r)+4 \mathcal{L}^2}{e^{2 (\lambda-\psi)} \left(m^2+r (r-2 m)\right)}\right]
\\
&&+\frac{1}{2} (\varepsilon^2-1)\ ,
\end{eqnarray}
where $e^\psi$ is evaluated at $\theta=\pi/2$. 
The radial dependence of the effective potential for various values of $\gamma$ and various values of the `quasi-NUT' parameter $p$ is shown in FIG.~\ref{veff_1}.
In the case $\gamma=1$ the effective potential takes the simple form
\bea 
V_{eff}^{\gamma=1}(r)=-\frac{1}{Fr^2}\left[m(pr-(1-p)m)-\frac{1}{2}\Delta L^2\right],
\eea 
and particle motion in the NUT space-time was studied in \cite{nut-particle}.

\begin{figure*}[t!]
\begin{center}
\includegraphics[width=0.45\linewidth]{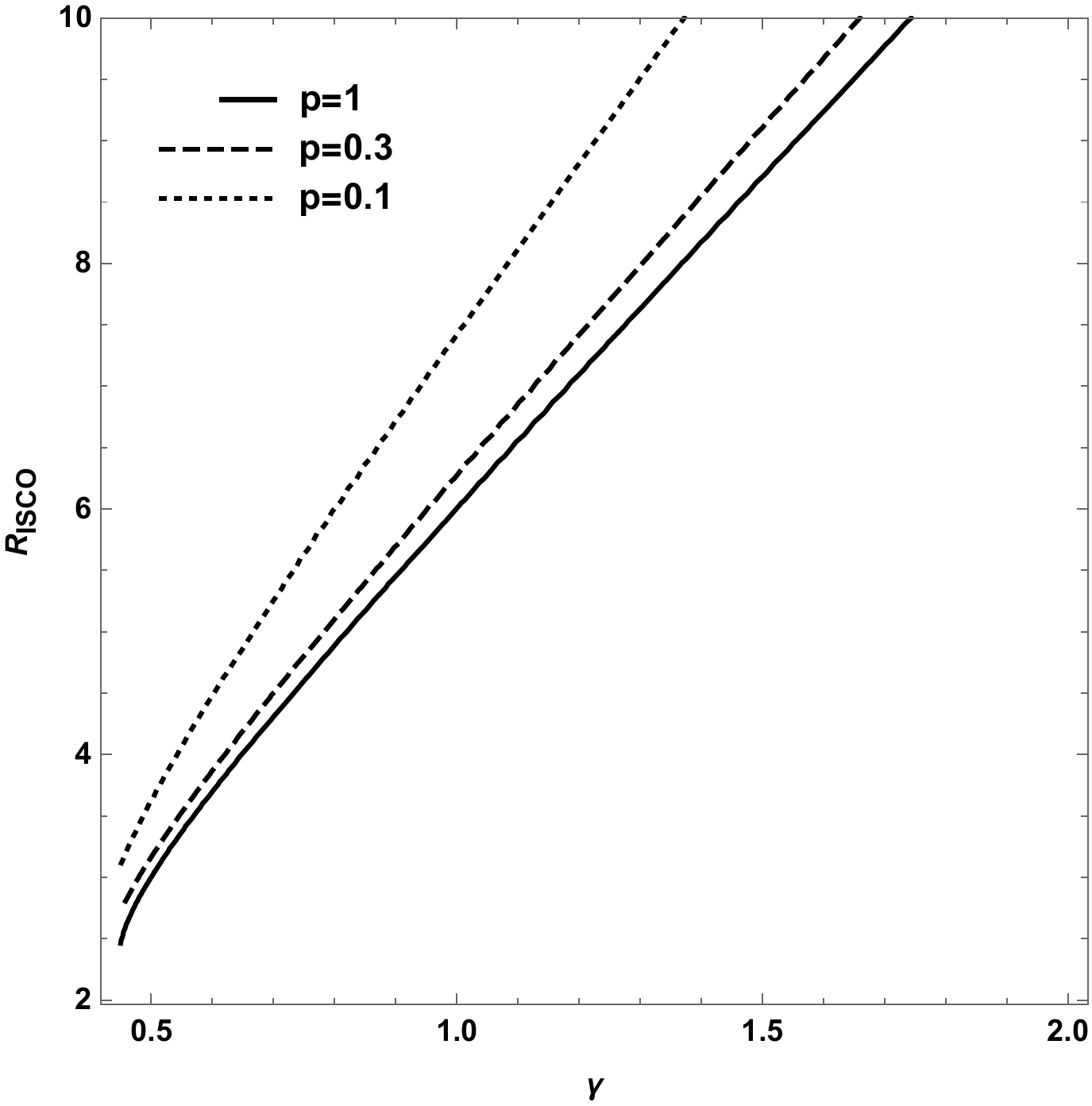}
\includegraphics[width=0.454\linewidth]{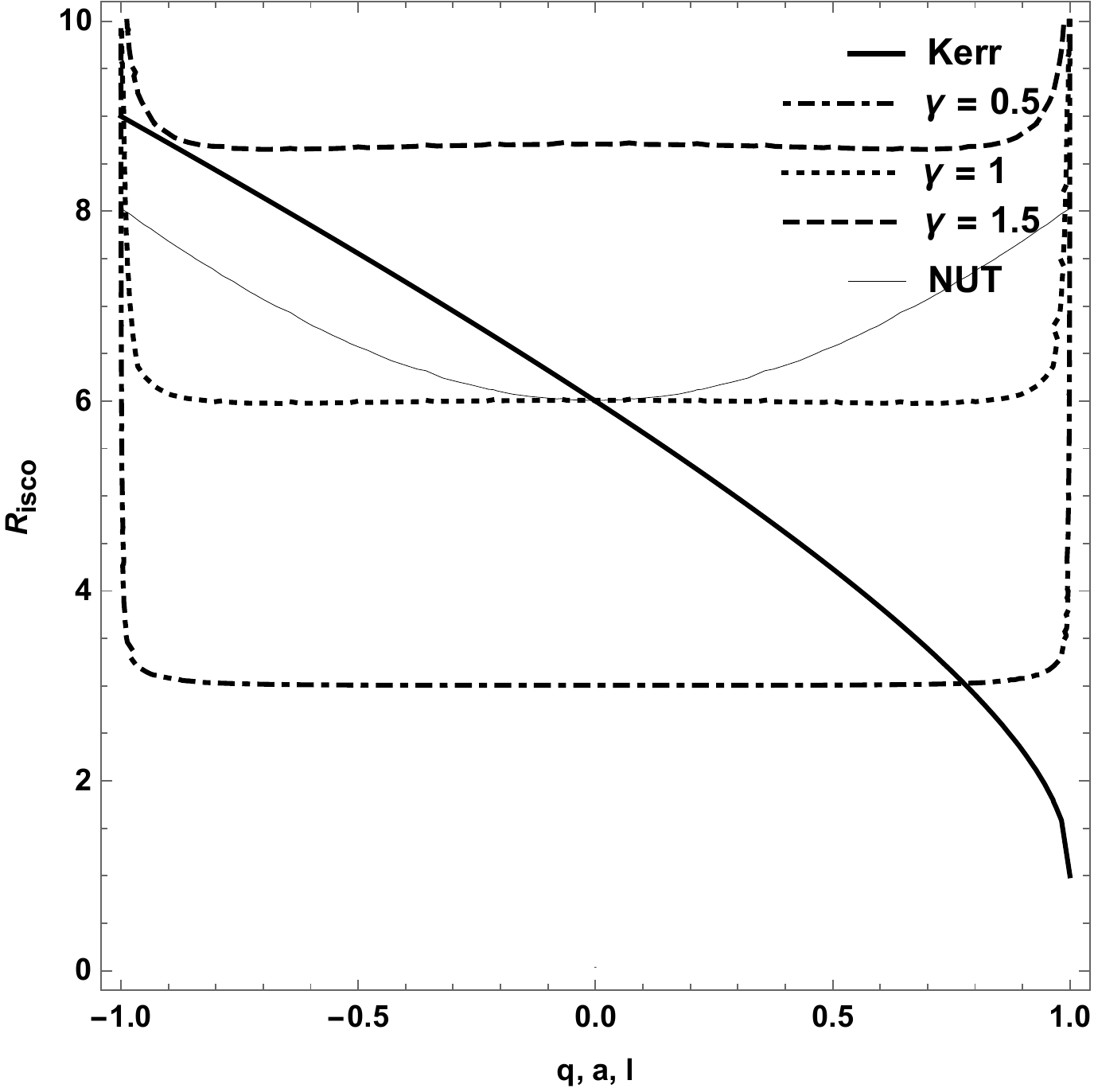}
\caption{The dependence of the ISCO radius of test particles on $\gamma$ and $q$. Left panel: The ISCO as a function of $\gamma$ for different values of $p=\sqrt{1-q^2}$. Right panel: The ISCO radius as a function of $q$ for different values of $\gamma$. For comparison we include the ISCO for Kerr and NUT. The parameter $a$ corresponds to the rotation parameter of the Kerr metric and $l$ corresponds to the gravitomagnetic charge of the NUT metric in the usual coordinates.
\label{isco_K}}
\end{center}
\end{figure*}

The innermost stable circular orbit (ISCO) of test particles in the equatorial plane can be calculated using the following standard conditions obtained from equation \eqref{f} \cite{nbm1}:
\begin{eqnarray}
f(r)&=&0 \ , \label{min}
\\
f'(r)&=&0 \ , \label{circle}
\\
f''(r)&=&0 \ . \label{isco}
\end{eqnarray}
Numerical values for the ISCO radius depending on $\gamma$ and $p$ are presented in TAB.\ref{1tab}.
From the table one can see that in the cases when $\gamma=1$ and $\gamma=1.5$ the ISCO has a minimum for a value of $p\in(0,1)$ (see right panel in Fig.~\ref{isco_K}). 
One can also compare the stationary ZV metric with Kerr and NUT in the usual coordinates (see Fig.~\ref{isco_K}). 
%It is interesting to notice how the actual parametrization affects the ISCO, since it is a coordinate dependent quantity. 
It is worth noticing that, even though the ISCO location is coordinate independent, its numerical value depends on the radial coordinate in use. In fact the case $\gamma=1$ and the NUT metric represent the same space-time in different coordinates and the two plots can be made to coincide making use of the change of coordinates \eqref{coord}.

\begin{table}
\caption{\label{1tab} The innermost stable circular orbits for test particles moving in the stationary ZV metric. }
\begin{ruledtabular}
\begin{tabular}{ccccccccc}
$p$ & 1 &0.8 & 0.6 & 0.3 & 0.1  \\
\hline
$\gamma=0.5$ & 3 & 3.0055 & 3.0291 & 3.1632 & 3.6186
\\
\hline
$\gamma=0.8$ & 4.888 & 4.859 & 4.887 & 5.119 & 5.981
\\
\hline
$\gamma=1$ & 6 & 5.9696 & 5.9858 & 6.2694 & 7.4137
\\
\hline
$\gamma=1.2$ & 7.097 & 7.059 & 7.081 & 7.425 & 8.811
\\
\hline
$\gamma=1.5$ & 8.7016 & 8.6505 & 8.6714 & 9.105 & 10.8755
\\
\end{tabular}
\end{ruledtabular}
\end{table}

It is worth remembering that the NUT space-time posses circular time-like geodesics outside the equatorial plane \cite{lam}. This can be seen also in the newly introduced coordinates for the NUT space-time given in equation \eqref{line-gamma=1}. In fact the components of the four velocity of a particle moving in the space-time described by the line element (\ref{line-gamma=1}) are
\begin{eqnarray}
\dot{t}&=&\frac{b(r) \varepsilon}{r (r-2 m)}-\frac{4 \varepsilon m^2 q^2 \cot ^2\theta -2 \mathcal{L}  m q \cot \theta  \csc \theta }{b(r)} ,
\\
\dot{\phi}&=&\frac{\csc ^2\theta  (2 \varepsilon m q \cos \theta +\cal{L})}{b(r)}\ ,
\\
\dot{r}&=&\frac{r(r-2 m)}{b(r)}  \sqrt{\frac{a(r) \varepsilon^2+b(r) +\mathcal{K}}{2 m r-r^2}}\ ,
\\
\dot{\theta}&=&\frac{\sqrt{\mathcal{K}-\csc ^2\theta  (2 \varepsilon m q \cos \theta +\mathcal{L})^2}}{b(r)}\ ,\label{thdot}
\end{eqnarray}
where $\mathcal{K}$ is a Carter's constant, $b(r)=r^2 - 2 m (1 - p) + 2 m^2 (1 - p)$, and $a(r)=-b(r)^2/(r^2-2mr)$. From equation (\ref{thdot}) one can show that there is off equatorial motion on a plane parallel to the equatorial one when
\begin{eqnarray}
\cos \theta = \frac{\pm\sqrt{4 \varepsilon^2 \mathcal{K} m^2 q^2+\mathcal{K}^2-\mathcal{K} \mathcal{L}^2}-2 \varepsilon \mathcal{L} m q}{4 \varepsilon^2 m^2 q^2+\mathcal{K}}.
\end{eqnarray}
Since we know that in the absence of parameter $q$ the metric must reduce to Schwarzschild we are led to choose Carter's constant as $\mathcal{K}=\mathcal{L}^2$. Then the above expression reduces to
\begin{eqnarray}
\cos \theta = \frac{-2 \varepsilon \mathcal{L} m q (1\pm1)}{4 \varepsilon^2 m^2 q^2+\mathcal{L}^2},
\end{eqnarray}
where the minus ($-$) sign gives us the motion on equatorial plane while the plus ($+$) sign gives the orbit confined to a cone with the opening angle $\theta$ given by 
\begin{eqnarray}
\cos \theta = \frac{-4 \varepsilon \mathcal{L} m q }{4 \varepsilon^2 m^2 q^2+\mathcal{L}^2}.
\end{eqnarray}

\begin{figure*}[t!]
\begin{center}
\includegraphics[width=0.49\linewidth]{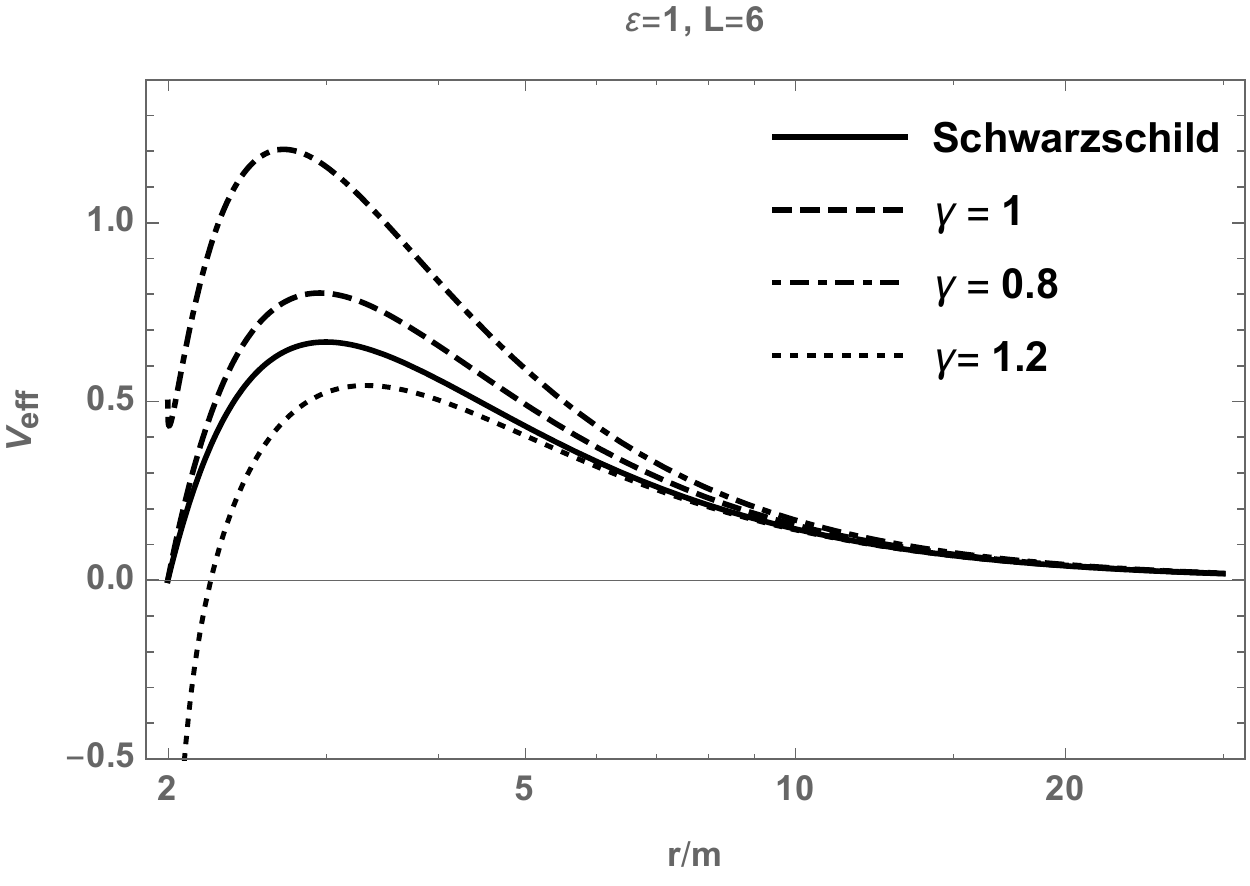}
\includegraphics[width=0.49\linewidth]{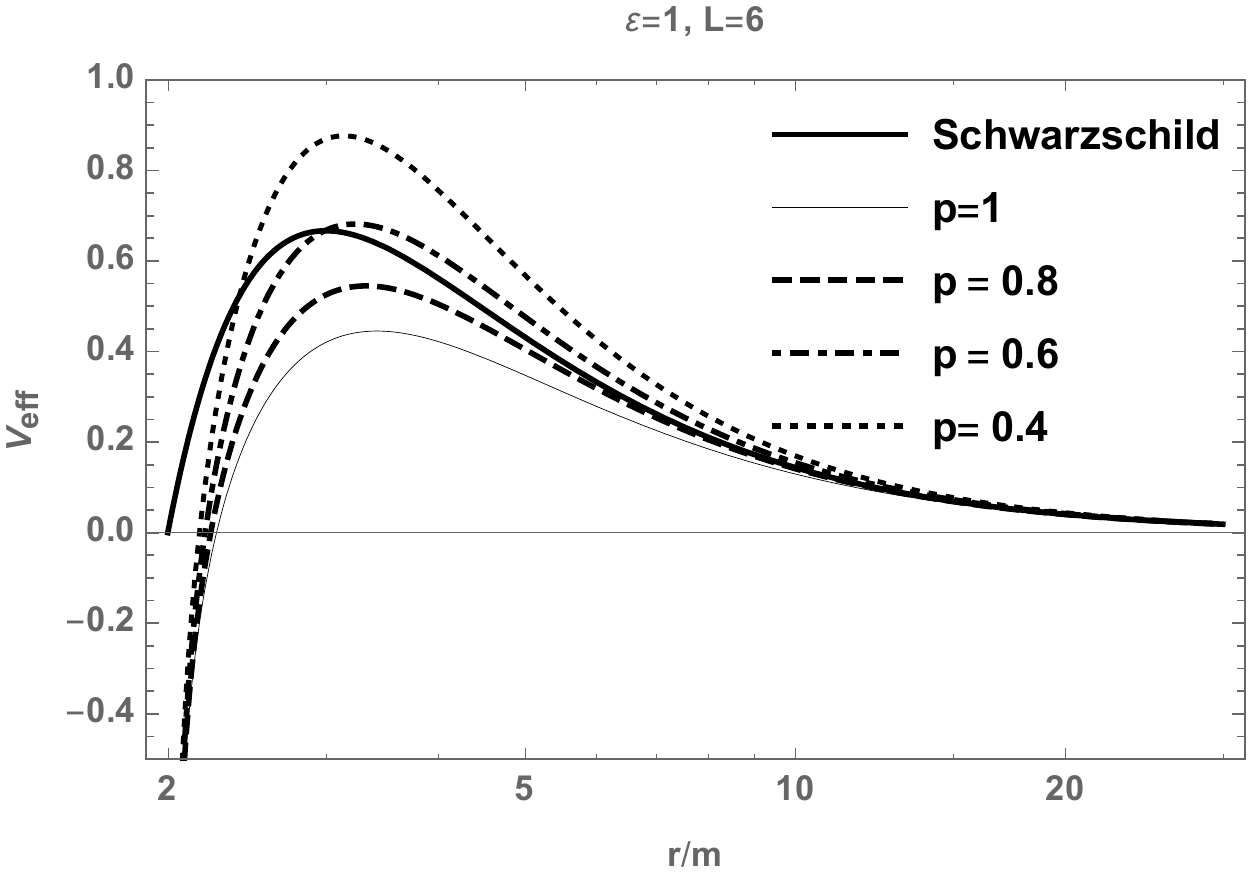}
\end{center}
\caption{The radial dependence of the effective potential $V_{eff}(r)$ for photons in the equatorial plane of the stationary ZV space-time as compared to the effective potential for Schwarzschild.  Left panel: $V_{eff}(r)$ is plotted for $p=0.8$ and various values of $\gamma$. Right panel: $V_{eff}(r)$ is plotted for $\gamma=1.2$ and various values of $p$. The effective potential for Schwarzschild (i.e. $\gamma=1$ and $p=1$) is included for reference. $R_{ISCO}$ is given in units of $m$.
\label{veff_ph}}
\end{figure*}

Let us now investigate the photon motion in the space-time characterized by the line element \eqref{line}. Using Hamilton-Jacobi equation of motion for massless particle one may derive the effective potential for photons as
\begin{eqnarray} \nonumber
&&V_{eff}= \left[\frac{\epsilon ^2}{2} + \frac{e^{-4 \psi} r \epsilon ^2 (2 m-r)+\csc ^2\theta ~(\mathcal{L}-\omega \epsilon )^2}{2e^{-4 \psi} e^{2 \lambda} \left(m^2+r \csc
   ^2\theta ~(r-2 m)\right) \sin ^2 \theta}\right]\ ,\\
   &&
\end{eqnarray}
where $\epsilon$ and $\mathcal{L}$ again define the conserved energy and angular momentum of photons. The radial dependence of this effective potential is plotted in Fig.~\ref{veff_ph}.  
One may check that for the effective potential of photons the sign of $V''_{eff}(r)$ is negative at points near the photon sphere showing that, as expected, photon orbits are unstable.

From the symmetry of the space-time one may calculate the radius of circular photon orbits on the equatorial plane ($\theta=\pi/2$) from either of the following geodesic equations
\begin{eqnarray}
ds^2=g_{\mu\nu}dx^{\mu} dx^{\nu}=0\ ,
\\
\ddot{x}^{\mu}+\Gamma^{\mu}_{\alpha\beta} \dot{x}^{\alpha}\dot{x}^{\beta}=0\ .
\end{eqnarray}
Setting $\ddot{r}=\dot{r}=\dot{\theta}=0$ we obtain following set of equations
\begin{eqnarray}
g_{tt}+2 g_{t\phi} \frac{d\phi}{dt}+g_{\phi\phi} \left(\frac{d\phi}{dt}\right)^2=0\ ,
\\
\Gamma^{r}_{tt}+2 \Gamma^{r}_{t\phi} \frac{d\phi}{dt}+\Gamma^{r}_{\phi\phi}\left(\frac{d\phi}{dt}\right)^2=0\ .
\end{eqnarray}

\begin{figure*}[t!]
\begin{center}
a.
\includegraphics[width=0.45\linewidth]{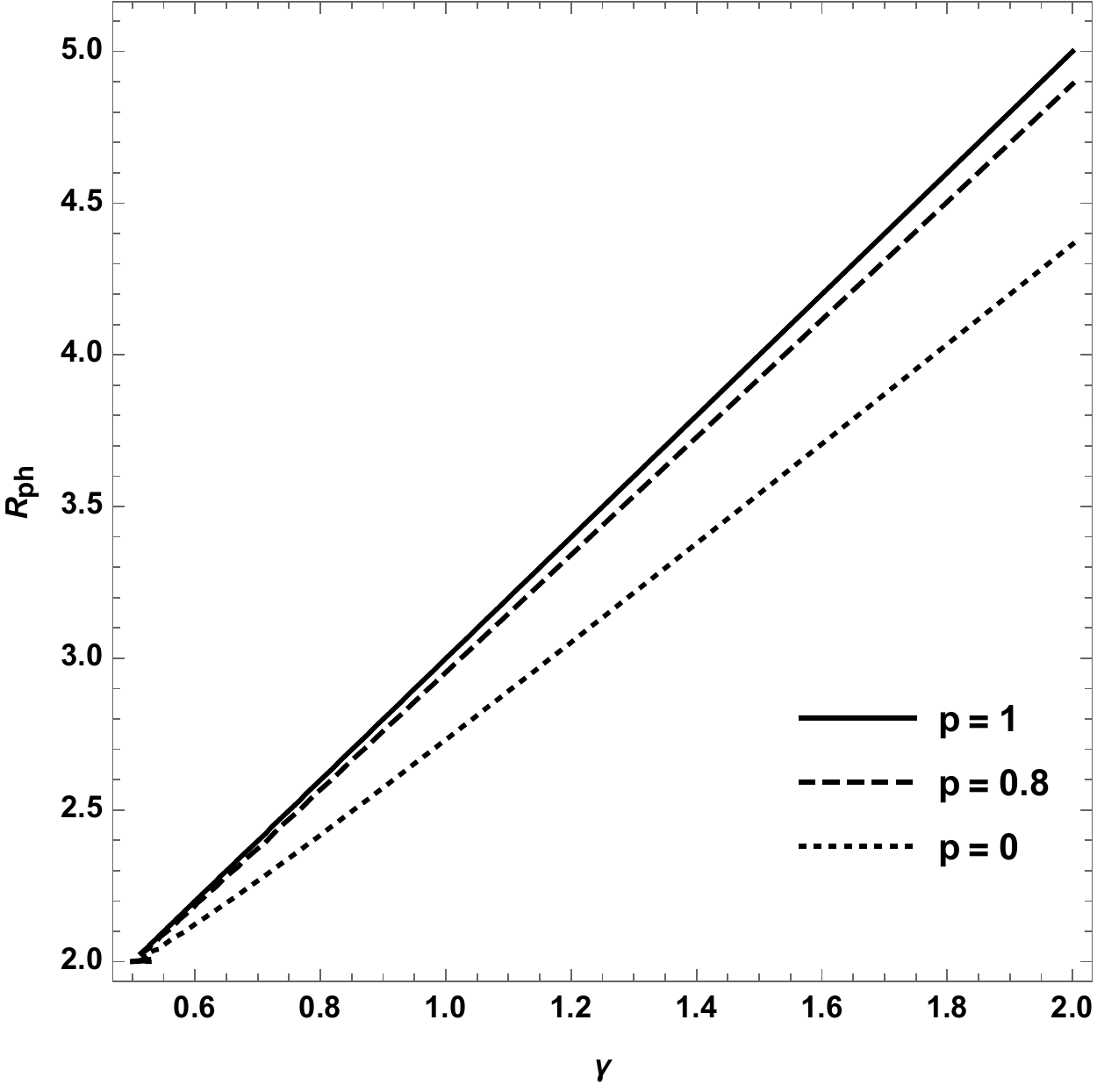}
b.
\includegraphics[width=0.45\linewidth]{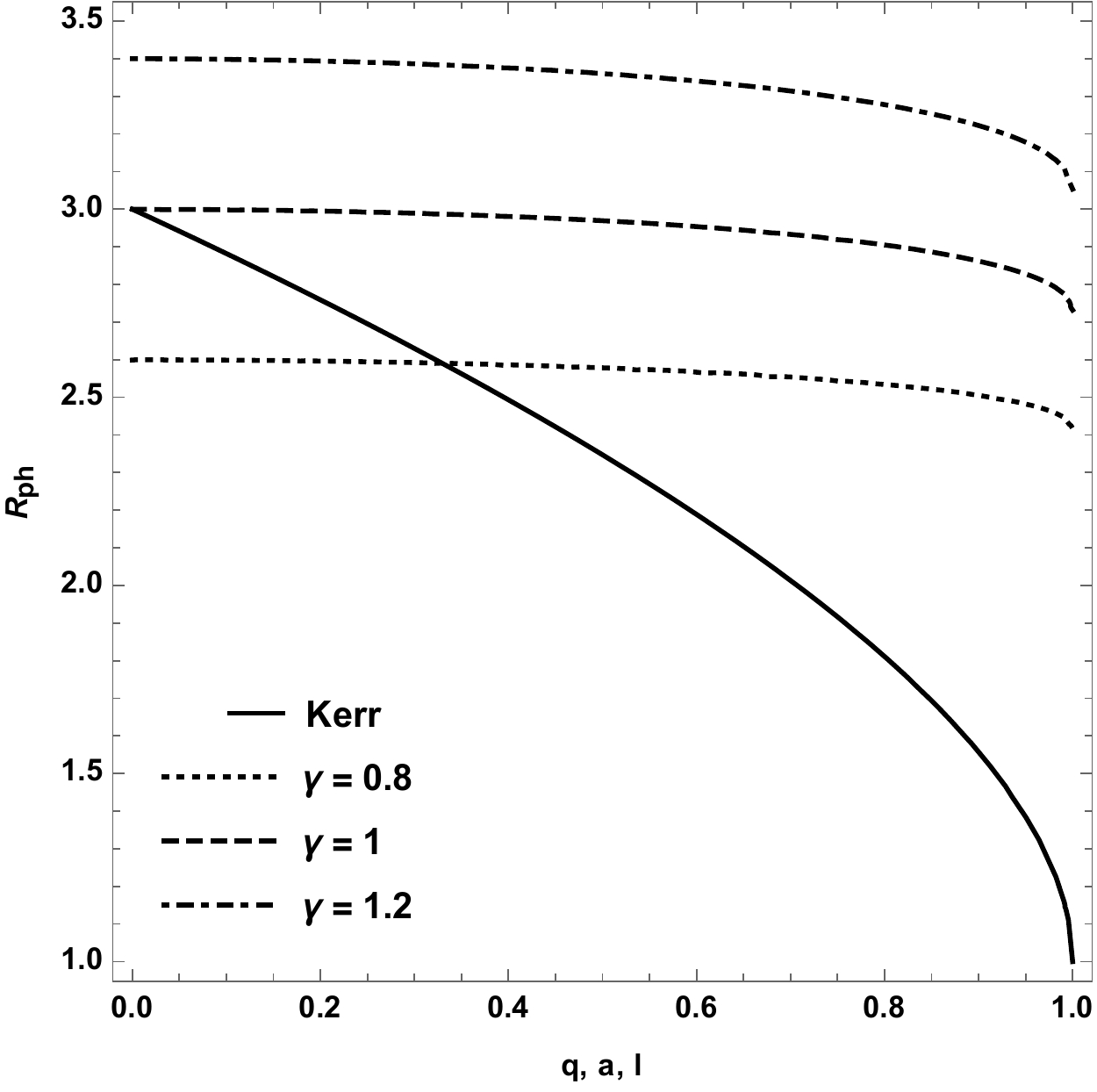}
\end{center}
\caption{Photon capture orbits in the stationary ZV space-time. Left panel: The dependence of the photon sphere radius on the deformation parameter $\gamma$ for different values of $q$. Right panel: The dependence of the photon sphere radius on the `quasi-NUT' parameter $q$ for different values of $\gamma$. For comparison we include the photon sphere for the Kerr. The parameter $a$ corresponds to the rotation parameter of the Kerr metric.
$R_{ph}$ is given in units of $m$. \label{psph}}
\end{figure*}

Solving the above equations analytically in the general case is complicated, however, some insights can be obtained from the plot of the photon capture radius in the equatorial plane as a function of $\gamma$ and $q$, as shown in Fig.~\ref{psph}.
It can be seen that the location of the photon capture radius increases with $\gamma$ similarly to what happens in the static case. Also the dependence of the photon sphere on the 'quasi-NUT' parameter shows how the space-time differs from Kerr. However, one needs to be careful with the interpretation of the radial coordinate as a radial distance, since a simple change of coordinate like the one in equation \eqref{coord} and a redefinition of the parameters like the ones given in equations \eqref{l} and \eqref{m} may completely change the behaviour of the photon sphere.

\begin{figure*}[hhh]
	\begin{center}
		\includegraphics[width=0.49\linewidth]{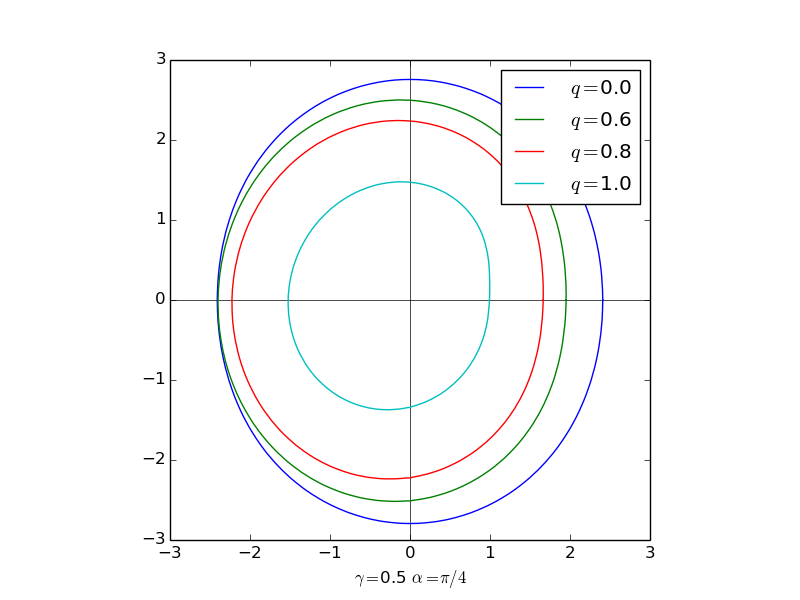}
		\includegraphics[width=0.49\linewidth]{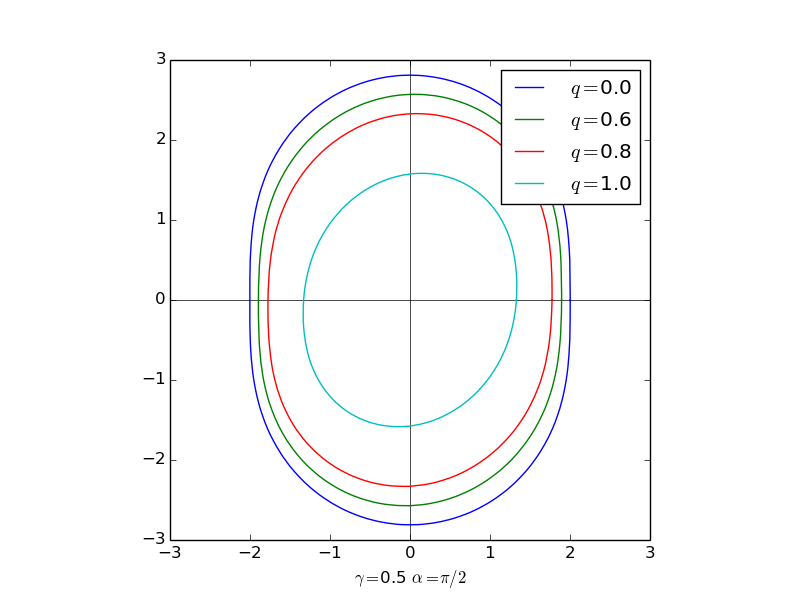}

		\includegraphics[width=0.49\linewidth]{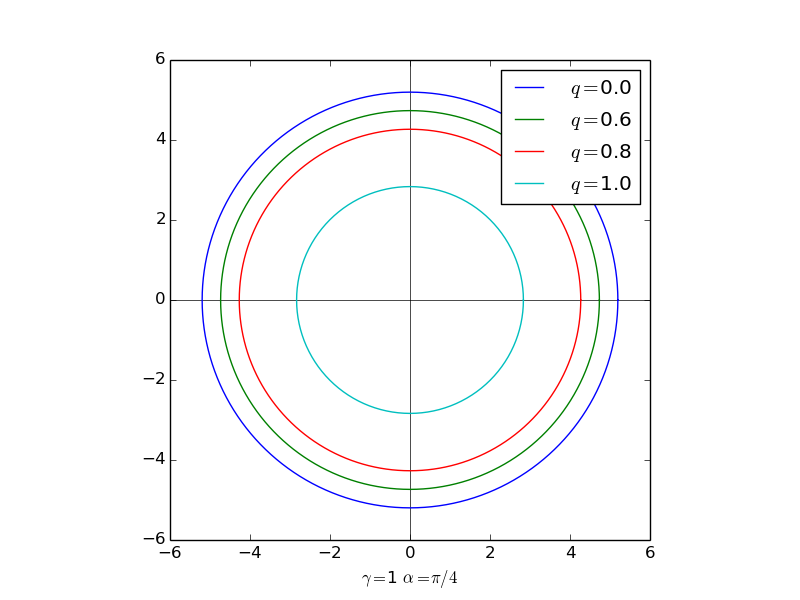}
		\includegraphics[width=0.49\linewidth]{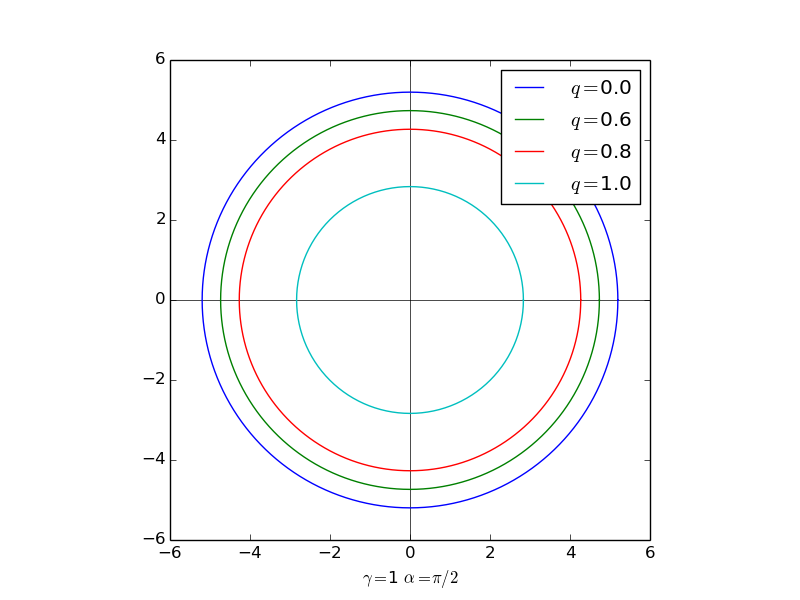}
		\end{center}
	\caption{Shadow of stationary ZV metric for $\gamma=1$ (NUT) and $\gamma=0.5$ at different inclination angles $\alpha$ for the observer and for different values of $q$. It can be seen that in the case $\gamma=1$ the introduction of the `quasi-NUT' parameter does not alter the shape of the shadow, thus showing again that $q$ must not be interpreted as a rotation parameter but is instead related to the NUT charge. On the other hand, departures from the NUT case, i.e. $\gamma\neq 1$ significantly alter the shape of the shadow, as expected.
		\label{shadow}}
\end{figure*}

Finally one may consider the appearance of the source for far away observers by evaluating the shape of the shadow of the line element \eqref{line}. The shadow of the Kerr-Taub-NUT black hole was considered in \cite{KTN} and the shadow of the static ZV metric was studied in \cite{shw1}. The shadow of the stationary ZV metric may be related to the shadow of the static ZV metric and that of the Kerr-NUT metric once the `quasi-NUT' parameter and the deformation parameter vanish. The shadow of the stationary ZV metric has been obtained using the ray tracing code that was developed in \cite{shw2, shw3, shw4, shw5} and it is shown in in Fig.~\ref{shadow}. It is clearly seen how the deformation parameter $\gamma$ and the `quasi-NUT' parameter affect the appearance of the source, thus suggesting that the nature geometry could in principle be tested via observations.

\section{Conclusion}\label{4}

We considered a stationary extension of the Zipoy-Voorhees space-time that was originally found by Halilsoy in \cite{halilsoy} in order to investigate if it could be taken as a suitable candidate to describe the field in the exterior of an astrophysical compact object or a black hole mimicker.

We found that the line element does not describe a rotating source but rather a deformed NUT space-time that reduces to the NUT metric in the limit of vanishing deformations, i.e. $\gamma=1$. Therefore the presence of the `quasi-NUT' parameter $q$ does not affect the motion of test particles in a way comparable to the Kerr parameter. In fact, in terms of the appearance of the source for far away observers the behaviour remains qualitatively similar to the ZV case, for all values of the `quasi-NUT' parameter.
Furthermore, for $\gamma\neq 1$, the metric exhibits the presence of closed time-like curves in a finite region of the space-time thus suggesting that the field may not represent the exterior of a viable source.

The effective potentials describing the motion of massive test particles and photons in the equatorial plane have been used to determine the location of the innermost stable circular orbit and the photon sphere. The dependence of these two radii on the deformation parameter and the `quasi-NUT' parameter shows that it would be in principle possible to observationally distinguish the Halilsoy geometry from Kerr and Schwarzschild.

\begin{acknowledgments}
	
The authors would like to thank Bobomurat Ahmedov and Naresh Dadhich, for useful comments and discussion and Askar Abdikamalov and Dimitry Ayzenberg for the code used to produce Fig.~\ref{shadow}.

This work was supported by the Innovation Program of the Shanghai Municipal Education Commission, Grant No.~2019-01-07-00-07-E00035, and the National Natural Science Foundation of China (NSFC), Grant No.~11973019. 
%and \textcolor{red}{ Fudan University (Grant No. IDH1512060)}.
%

This research is supported by Grants No. VA-FA-F-2-008 and No. MRB-AN-2019-29 of the Uzbekistan Ministry for Innovative Development.

B.N. also acknowledges support from the China Scholarship Council (CSC), grant No.~2018DFH009013.

B.N. and D.M. acknowledge support from Nazarbayev University Faculty Development Competitive Research Grant No. 090118FD5348.
	
\end{acknowledgments}

\end{document}